 \documentclass[manuscript]{emulateapj-rtx4}

\newcommand{\myemail}{hirotani@tiara.sinica.edu.tw}

\slugcomment{\it Received 2012 November 19; accepted 2013 January 22}

\shorttitle{Luminosity Evolution of Gamma-ray Pulsars}
\shortauthors{Hirotani}

\begin{document}


\title{Luminosity Evolution of Gamma-ray Pulsars}


\author{Kouichi Hirotani\altaffilmark{1}}
\affil{Theoretical Institute for
       Advanced Research in Astrophysics (TIARA),
       Academia Sinica, Institute of Astronomy and Astrophysics (ASIAA),
       PO Box 23-141, Taipei, Taiwan}
\myemail


\altaffiltext{1}{Postal address: 
                 TIARA, Department of Physics, 
                 National Tsing Hua University,
                 101, Sec. 2, Kuang Fu Rd.,Hsinchu, Taiwan 300}


\begin{abstract}
We investigate the electrodynamic structure of
a pulsar outer-magnetospheric particle accelerator
and the resultant gamma-ray emission.
By considering the condition for the
accelerator to be self-sustained,
we derive how the trans-magnetic-field thickness of the accelerator
evolves with the pulsar age.
It is found that the thickness is small but increases steadily
if the neutron-star envelope is contaminated by sufficient light elements.
For such a light element envelope,
the gamma-ray luminosity of the accelerator 
is kept approximately constant as a function of age
in the initial ten thousand years,
forming the lower bound of the observed distribution
of the gamma-ray luminosity of rotation-powered pulsars.
If the envelope consists of only heavy elements, on the other hand,
the thickness is greater but increases less rapidly
than what a light element envelope has.
For such a heavy element envelope,
the gamma-ray luminosity decreases relatively rapidly,
forming the upper bound of the observed distribution.
The gamma-ray luminosity of a general pulsar 
resides between these two extreme cases, 
reflecting the envelope composition and the magnetic inclination angle
with respect to the rotation axis.
The cutoff energy of the primary curvature emission is
regulated below several GeV even for young pulsars,
because the gap thickness, and hence the acceleration electric field
is suppressed by the polarization of the produced pairs.
\end{abstract}





\section{Introduction}
The Large Area Telescope aboard {\it Fermi Gamma-ray Space Telescope} 
\citep{atwood09} has proved remarkably successful 
at discovering rotation-powered pulsars emitting photons above
0.1~GeV.
Thanks to its superb sensitivity,
the number of gamma-ray pulsars has increased from
six in Compton Gamma Ray Observatory era \citep{thomp04}
to more than one hundred \citep{nolan12}. 
Plotting their best estimate of
the gamma-ray luminosity, $L_\gamma$, 
against the spin-down luminosity,
$L_{\rm spin}=4\pi^2 I \dot{P} P^{-3}$,
\citet{abd10} found the important relation
that $L_\gamma$ is approximately proportional to 
$L_{\rm spin}{}^{0.5}$ (with a large scatter), 
where $I$ refers to the neutron-star (NS) moment of inertia,
$P$ the NS rotational period, and 
$\dot{P}$ its temporal derivative.
However, it is unclear why this relationship arises, 
in spite of its potential importance to discriminate
pulsar emission models such as 
the polar-cap model \citep{harding78,daugherty82,dermer94},
the outer-gap model \citep{chiang92,romani96,zhang97,tak06,hiro08,wangY11},
the pair-starved polar-cap model \citep{vent09}
(see also \citet{yuki12} for the possible co-existence of such models),
and the emission model 
from the wind zone \citep{petri10,bai10a,bai10b,aha12}.

Recent gamma-ray observations 
suggest that the pulsed gamma-rays are emitted from the
higher altitudes of a pulsar magnetosphere.
This is because the observed light curves \citep{abd10} favor 
fan-like emission geometry, 
which scan over a large fraction of the celestial sphere,
and because the Crab pulsar shows pulsed photons 
near and above 100~GeV \citep{aliu11,ale11a,ale11b},
which rules out an emission from the lower altitudes,
where strong magnetic absorption takes place for $\gamma$-rays
above $10$~GeV.
Consequently, higher-altitude emission models
such as the outer-gap model \citep{cheng86a,cheng86b},
the high-altitude slot-gap model \citep{musl04},
or the pair-starved polar-cap model \citep{vent09},
gathered attention.
It is noteworthy that the outer-gap model is presently 
the only higher-altitude emission model that is solvable 
from the basic equations self-consistently \citep{hiro11a}. 
In the present paper, therefore, 
we focus on the outer-gap model and derive the
observed relationship $L_\gamma \propto L_{\rm spin}{}^{0.5}$
both analytically and numerically.

\begin{figure}
\epsscale{1.0}
\plotone{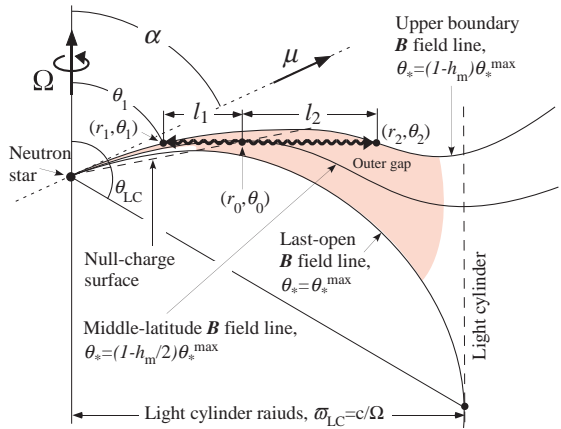}  
\caption{
Side view of an outer gap.
The neutron star (filled circle on the left) 
obliquely rotates around the vertical axis 
with magnetic inclination angle $\alpha$.
The thin solid curves denote the magnetic field lines,
while the dashed straight line the null-charge surface,
on which the magnetic field lines become perpendicular to the
rotation axis.
Outside the light cylinder (the vertical long dashed line),
plasmas that are frozen to the magnetic field lines
can only migrate outwards (as a pulsar wind) 
because of the causality requirement in special relativity.
The light-cylinder radius, $\varpi_{\rm LC}$, becomes 
typically a few or several hundred neutron-star radii
for young pulsars.
In modern outer-gap models \citep{hir03,tak04,hir06a,tak08}, 
it is proved that
the outer gap extends between the stellar surface
(because of a negative charge density in the lower altitudes)
and the vicinity of the light cylinder
(because of a positive charge density in the higher altitudes).
Thus, typical inward (or outward) photons propagate
distance $l_1$ (or $l_2$) before escaping from the gap
(shaded region).
\label{fig:sideview}
}
\end{figure}

We schematically depict 
the pulsar outer-magnetospheric accelerator (i.e., the outer gap)
in figure~\ref{fig:sideview}.
As the NS rotates, there appears the light cylinder,
within which plasmas can co-rotate with the magnetosphere.
The magnetic field lines that become tangential to the
light cylinder at the light cylinder radius,
$\varpi_{\rm LC}=cP/2\pi$, 
are called the last-open magnetic field lines,
where $c$ refers to the speed of light.
Pairs are produced via photon-photon pair production 
mostly near the null-charge surface 
and quickly polarized by 
the magnetic-field-aligned electric field, $E_\parallel$, in the gap. 
In this paper, we assume that the
rotation and magnetic axes reside in the same hemisphere
to obtain $E_\parallel > 0$,
which accelerates positrons ($e^+$'s) outwards while
electrons ($e^-$'s) inwards.
These ultra-relativistic particles 
have Lorentz factors, $\gamma \sim 10^{7.5}$,
to emit photons efficiently by the curvature process.

\section{Analytical examination of outer-gap luminosity}
\label{sec:analytical}
In this section, we analytically derive 
how the gamma-ray luminosity of an outer gap evolves with time.
In the outer magnetosphere, only the dipole component remains
in the magnetic field;
thus, the inhomogeneous part of the Maxwell equation 
(i.e., the Poisson equation for the electro-static potential)
gives the magnetic-field-aligned electric field \citep{hiro08}, 
\begin{equation}
  E_\parallel \approx \frac{\mu}{2\varpi_{\rm LC}^3} h_{\rm m}^2,
  \label{eq:app_Ell}
\end{equation}
where $\mu$ denotes the NS magnetic dipole moment,
and $h_{\rm m}$ the trans-magnetic-field thickness of the gap.
Since the Poisson equation is a second-order differential equation,
$E_\parallel$ is proportional to $h_{\rm m}{}^2$.
Electrons ($e^-$'s) and positrons ($e^+$'s) 
are created via photon-photon 
(and sometimes via magnetic) pair production, 
being subsequently polarized by $E_\parallel$ 
and accelerated in the opposite directions,
to finally attain the terminal Lorentz factor 
\begin{equation}
  \gamma = \left( \frac{3 \rho_{\rm c}^2}{2e} E_\parallel 
           \right)^{1/4},
  \label{eq:app_Lf}
\end{equation}
where $\rho_{\rm c}$ refers to the radius of curvature of
particle's motion in the three-dimensional magnetosphere,
$e$ the charge on the positron.
Photons are radiated by such ultra-relativistic
$e^\pm$'s via curvature process with characteristic energy,
\begin{equation}
  h \nu_{\rm c}= \frac32 \hbar c \frac{\gamma^3}{\rho_{\rm c}},
  \label{eq:app_Eg}
\end{equation}
where $h$ denotes the Planck constant,
$\hbar \equiv h/2\pi$.
Once $h_{\rm m}$ is obtained, 
we can readily compute the $\gamma$-ray luminosity
of curvature radiation from an outer gap by \citep{hiro08}
\begin{equation}
  L_\gamma \approx 2.36 (\nu F_\nu)_{\rm peak} \times 4\pi d^2 f_\Omega
          \approx 1.23 f_\Omega h_{\rm m}^3 
                  \frac{\mu^2 \Omega^4}{c^3},
  \label{eq:Lg_0}
\end{equation}
where $f_\Omega$, 
which has been conventionally assumed to be approximately unity, 
refers to the flux correction factor \citep{romani10},
and $\Omega=2\pi/P$ the rotation angular frequency of the NS.
Here, it is assumed that the current density flowing in the gap
is comparable to the Goldreich-Julain value \citep{gol69}.
The last factor, $\mu^2 \Omega^4/c^3$ is proportional to the spin-down 
luminosity, $L_{\rm spin}$.
Therefore, the evolution law, $L_\gamma \propto L_{\rm spin}^{0.5}$, 
is crucially governed by the evolution of $h_{\rm m}$
as a function of the NS age, $t$.

The evolution of $h_{\rm m}$ is essentially controlled by
the photon-photon pair production in the pulsar magnetosphere.
To analytically examine the pair production,
we assume the static dipole magnetic field configuration
for simplicity,
and consider the plane on which both the rotational and magnetic
axes reside (fig.~\ref{fig:sideview}).
On this two-dimensional latitudinal plane, 
the last-open field line intersects the NS surface
at magnetic co-latitudinal angle $\theta_\ast^{\rm max}$
(measured from the magnetic dipole axis) that satisfies
\begin{equation}
  \frac{\sin^2\theta_\ast^{\rm max}}{r_\ast}
  = \frac{\sin^2(\theta_{\rm LC}-\alpha)}
         {\varpi_{\rm LC}/\sin\theta_{\rm LC}},
\end{equation}
where $r_\ast$ denotes the NS radius,
$\theta_{\rm LC}$ the angle 
(measured from the rotation axis)
of the point where the last-open field line becomes
tangential to the light cylinder,
and $\alpha$ the inclination angle of the dipole magnetic axis
with respect to the rotation axis.
A magnetic field line can be specified by the magnetic
co-latitude (measured from the dipole axis), 
$\theta_\ast$,
where it intersects the stellar surface.
A magnetic field line does not close within the light cylinder
(i.e., open to large distances)
if $0<\theta_\ast<\theta_\ast^{\rm max}$.
Thus, the last-open field lines, $\theta_\ast = \theta_\ast^{\rm max}$, 
corresponds to the {\it lower} boundary, 
which forms a surface in a three-dimensional magnetosphere, 
of the outer gap.

Let us assume that the gap {\it upper} boundary 
coincides with the magnetic field lines that are specified by
$\theta_\ast= (1-h_{\rm m})\theta_\ast^{\rm max}$.
Numerical examinations show
that $h_{\rm m}$, indeed, changes as a function of the distance
along the field line and the magnetic azimuthal angle
(measured counter-clockwise around the dipole axis).
Nevertheless, except for young pulsars like the Crab pulsar,
an assumption of a spatially constant $h_{\rm m}$ gives
a relatively good estimate.
Thus, for an analytical purpose,
we adopt a constant $h_{\rm m}$ in this analytical examination.
In this case, we can specify
the middle-latitude field line by the magnetic co-latitude
$\theta_\ast= (1-h_{\rm m}/2)\theta_\ast^{\rm max}$.
Screening of $E_\parallel$ due to the polarization of the
produced pairs,
takes place mostly in the lower altitudes.
It is, therefore, appropriate to evaluate the screening of $E_\parallel$
around the point ($r_0$,$\theta_0$)
where the null-charge surface intersects the middle-latitude field line
(fig.~\ref{fig:sideview}).

An inwardly migrating electron
(or an outwardly migrating positron)
emits photons inwards (or outwards),
which propagate the typical distance $l_1$ (or $l_2$)
before escaping from the gap.
Denoting the cross section of the inward (or outward) horizontal line
from the point ($r_0$,$\theta_0$) and the
upper boundary as ($r_1$,$\theta_1$)
(or as ($r_2$,$\theta_2$)),
and noting 
$r_0 \cos\theta_0=r_1\cos\theta_1=r_2\cos\theta_2$, 
we obtain
\begin{equation}
  l_1= r_0 \cos\theta_0 (\tan\theta_0 -\tan\theta_1),
  \label{eq:S_l1}
\end{equation}
\begin{equation}
  l_2= r_0 \cos\theta_0 (\tan\theta_2 -\tan\theta_0).
  \label{eq:S_l2}
\end{equation}
Along the upper-boundary field line, we obtain
\begin{equation}
    \frac{\sin^2(\theta_1-\alpha)}{r_1}
  = \frac{\sin^2(\theta_2-\alpha)}{r_2}
  = \frac{\sin^2[(1-h_{\rm m})\theta_\ast^{\rm max}]}
         {r_\ast},
\end{equation}
whereas along the middle-latitude field line, we obtain
\begin{equation}
  \frac{\sin^2(\theta_0-\alpha)}{r_0}
  = \frac{\sin^2[(1-h_{\rm m}/2)\theta_\ast^{\rm max}]}
         {r_\ast}.
\end{equation}
Combining these two equations,
and noting $\theta_\ast^{\rm max} \ll 1$,
we find that $\theta_1$ ($<\theta_0$) and 
$\theta_2$ ($>\theta_0$)
can be given by the solution $\theta$ that satisfies
\begin{equation} 
  \cos\theta\sin^2(\theta-\alpha)
  = \left(\frac{1-h_{\rm m}}{1-h_{\rm m}/2}\right)^2
    \cos\theta_0 \sin^2(\theta_0-\alpha),
  \label{eq:S_th1}
\end{equation}
where $\theta_0$ is given by
\begin{equation} 
  \tan\theta_0
  = \frac12 
    \left( 3\tan\alpha+\sqrt{9\tan^2\alpha+8} \right). 
  \label{eq:S_th0}
\end{equation}
Thus, if we specify $\alpha$, 
we can solve $\theta=\theta_1$ and $\theta=\theta_2$ 
as a function of $h_{\rm m}$ by equation~(\ref{eq:S_th1}).
Substituting these $\theta_1$ and $\theta_2$ into
equations~(\ref{eq:S_l1}) and (\ref{eq:S_l2}),
we obtain $l_1$ and $l_2$,
where $r_0$ depends on $P$.

If $h_{\rm m} \ll 1$, we can expand the left-hand side of
equation~(\ref{eq:S_th1}) around
$\theta=\theta_0$,
where $\theta=\theta_1$ for inward 
(or $\theta=\theta_2$ for outward) $\gamma$-rays
to find $\theta_2-\theta_0= \theta_0-\theta_1 \propto \sqrt{h_{\rm m}}$.
That is, the leading terms in the expansion vanish and we obtain
$l_1= l_2$ from the next-order terms,
which are quadratic to $\theta-\theta_0$.
Assuming $L_{\rm X}\propto t^{-\beta}$,
where $\beta \approx 0.48$ is appropriate
for $t<10^4$ years for a light-element-envelope NS
and for $t<10^5$ years for a heavy-element NS,
we find
$h_{\rm m} \propto P^{5/6} \mu^{-1/6} t^{\beta/2}$,
and hence $L_\gamma \propto P^{-3/2} \mu^{3/2} t^{3\beta/2}$.
Since the dipole radiation formula gives
$P \propto \mu t^{1/2}$, we obtain
$L_\gamma \propto \mu^0 t^{3(\beta-1/2)/2} \propto t^{-0.03}$.
Thus, when the gap is very thin,
which is expected for a light-element-envelope NS,
$L_\gamma$ little evolves with the pulsar age, $t$.

However, if $h_{\rm m}>0.2$, say,
the rapidly expanding magnetic flux tube 
gives asymmetric solution, $l_2 > l_1$.
That is, the third and higher order terms in the expansion contribute
significantly compared to the quadratic terms.
Thus, we must solve equation~(\ref{eq:S_th1})
for $\theta$ ($=\theta_1$ or $\theta_2$)
without assuming $\vert \theta-\theta_0 \vert \ll 1$, in general.

Let us now consider the condition for a gap to be self-sustained.
A single ingoing $e^-$ or an outgoing $e^+$ emits
\begin{equation}
  (N_\gamma)_1= e E_\parallel l_1 / (h \nu_{\rm c})
\end{equation}
or
\begin{equation}
  (N_\gamma)_2= e E_\parallel l_2 / (h \nu_{\rm c})
\end{equation}
photons while running the typical distance $l_1$ or $l_2$, respectively.
Such photons materialize as pairs with probability
\begin{equation}
  \tau_1= l_1 F_1 \sigma_1 / c
\end{equation}
or 
\begin{equation}
  \tau_2= l_2 F_2 \sigma_2 / c,
\end{equation}
where $F_1$ and $F_2$ denote the X-ray flux 
inside and outside of ($r_0$,$\theta_0$), respectively; 
$\sigma_1$ and $\sigma_1$ are the pair-production cross section
for inward and outward photons, respectively.
Thus, a single $e^-$ or $e^+$ cascades into 
\begin{equation}
  (N_\gamma)_1 \tau_1= \frac{e E_\parallel}{h \nu_{\rm c}}
                      \frac{F_1}{c}
                      l_1{}^2 \sigma_1
  \label{eq:S_N1}
\end{equation}
pairs or into
\begin{equation}
  (N_\gamma)_2 \tau_2= \frac{e E_\parallel}{h \nu_{\rm c}}
                      \frac{F_2}{c}
                      l_2{}^2 \sigma_2
  \label{eq:S_N2}
\end{equation}
pairs within the gap.
That is, a single inward-migrating $e^-$ cascades into 
pairs with multiplicity $(N_\gamma)_1 \tau_1$.
Such produced pairs are polarized by $E_\parallel$.
Each returning, outward-migrating $e^+$ cascades into
pairs with multiplicity $(N_\gamma)_2 \tau_2$ in outer magnetosphere.
As a result, a single inward $e^-$ cascades eventually
into $(N_\gamma)_1 \tau_1 \cdot (N_\gamma)_2 \tau_2$ inward $e^-$'s,
which should become unity for the gap to be self-sustained.
Therefore, in a stationary gap,
the gap thickness $h_{\rm m}$ is automatically regulated
so that the gap closure condition,
\begin{equation}
  (N_\gamma)_1 \tau_1 \cdot (N_\gamma)_2 \tau_2 = 1,
  \label{eq:S_stationary}
\end{equation}
may be satisfied.

Approximately speaking, a single, inward-migrating
$e^-$ emits $(N_\gamma)_1 \sim 10^4$ curvature photons,
a portion of which head-on collide the surface X-ray photons
to materialize as pairs with probability
$\tau_1 \sim 10^{-3}$ within the gap.  
Thus, a single $e^-$ cascades into 
$(N_\gamma)_1 \tau_1 \sim 10$ pairs in the gap.
Each produced $e^+$ return outwards to emit 
$(N_\gamma)_2 \sim 10^5$ photons,
which materialize as pairs with probability
$\tau_2 \sim 10^{-6}$ 
by tail-on colliding with the surface X-rays.
In another word, $(N_\gamma)_1 \tau_1\sim 10$ 
holds regardless of the nature of the pair production process
(e.g., either photon-photon or magnetic process \citep{tak10})
in the lower altitudes,
because 
it is determined by the pair-production efficiency
in the outer magnetosphere
$(N_\gamma)_2 \tau_2\sim 0.1$,
which is always due to photon-photon pair production.

In general, 
$(N_\gamma)_1$, $\tau_1$, $(N_\gamma)_2$, $\tau_2$
are expressed in terms of 
$h_{\rm m}$, $P$, $\mu$, $T$, and $\alpha$,
where $T$ denotes the NS surface temperature.
Note that we can solve $P=2\pi/\Omega$ as a function of 
the NS age, $t$,
from the spin-down law.
Thus, specifying $\alpha$ and the cooling curve, $T=T(t)$,
we can solve $h_{\rm m}$ as a function of $t$
from the gap closure condition,
$(N_\gamma)_1 \tau_1 (N_\gamma)_2 \tau_2 = 1$.
Note also that the spin-down law
readily gives the spin-down luminosity,
$L_{\rm spin} \propto \dot{P} P^{-3}$,
as a function of $t$, 
once $P=P(t,\alpha)$ is solved.
On these grounds,
$L_\gamma$ can be related to $L_{\rm spin}$
with an intermediate parameter $t$,
if we specify the cooling curve and the spin-down law.

Substituting equations~(\ref{eq:S_N1}) and (\ref{eq:S_N2})
into (\ref{eq:S_stationary}), 
we obtain 
\begin{equation}
  \frac{e E_\parallel}{h\nu_{\rm c}}
  \frac{\sqrt{F_1 \sigma_1 F_2 \sigma_2}}{c}
  l_1 l_2 = 1,
  \label{eq:S_master}
\end{equation}
where 
\begin{equation}
  F_i \sigma_i
  = \pi (1-\mu_i) \left(\frac{r_\ast}{r_i}\right)^2
    \int_{\nu_{{\rm th},i}}^\infty \frac{B_\nu(T)}{h\nu} 
                      \sigma_{\rm P}(\nu,\nu_\gamma,\mu_i)
\end{equation}
with $i=1,2$; $\nu_\gamma$ denotes
the $\gamma$-ray frequency, and $B_\nu(T)$ the Planck function.
We have to integrate over the soft photon frequency
$\nu$ above the threshold energy
\begin{equation}
  h\nu_{{\rm th},i}= \frac{2(m_{\rm e}c^2)^2}
               {(1-\mu_i)h\nu_\gamma},
\end{equation}
where $m_{\rm e}c^2$ refers to the rest-mass energy of the electron.
The cosine of the collision angle $\mu_i$ becomes
$1-\mu_1 = 1 - \sin\theta_0$ for outward 
(or $1-\mu_2 = 1+\sin\theta_0$ for inward) $\gamma$-rays.
That is, collisions take place head-on (or tail-on)
for inward (or outward) $\gamma$-rays.
The total cross section is given by
\begin{equation}
  \sigma_{\rm P}
  =\frac{3}{16} \sigma_{\rm T} 
   (1-v^2)
   \left[ (3-v^4) \ln\frac{1+v}{1-v} -2v(2-v^2) \right],
\end{equation}
where $\sigma_{\rm T}$ denotes the Thomson cross section and
\begin{equation}
  v \equiv \sqrt{1-\frac{2}{1-\mu_i}
                   \frac{(m_{\rm e}c^2)^2}{h\nu h\nu_\gamma}}.
\end{equation}

Pair production takes place when the $\gamma$-rays collide with the
surface X-rays in the Wien regime,
that is, at $h\nu \gg kT$.
An accurate evaluation of $\sigma_2$ requires
a careful treatment of the collision geometry,
because the threshold energy, $h\nu_{{\rm th},2}$,
strongly depends on the tiny collision angles.
In the numerical method (next section),
the pair-production absorption coefficient is explicitly computed
at each point in the three-dimensional pulsar magnetosphere
by equation~(\ref{eq:def_alf2}).
However, in this section, for analytical purpose,
we simply adopt the empirical relation,
\begin{equation}
  \sqrt{F_1 \sigma_1 F_2 \sigma_2}
  = \epsilon \sqrt{1-\mu_1} \sigma_{\rm T} F_{\rm X},
  \label{eq:L1L2}
\end{equation}
where $\epsilon=0.004$, $0.01$, and $0.038$ for
$\alpha=45^\circ$, $60^\circ$, and $75^\circ$, respectively;
$1-\mu_1 \approx 2$.
The X-ray flux is evaluated at ($r_0$,$\theta_0$) such that
\begin{equation}
  F_{\rm X} = \frac{L_{\rm X}}{2.82 kT}
             \frac{1}{4\pi r_0{}^2},
\end{equation}
where $L_{\rm X}$ refers to the luminosity of photon radiation 
from the the cooling NS surface.
For a smaller $\alpha$, the point ($r_2$,$\theta_2$) is located
in the higher altitudes,
where the magnetic field lines begin to collimate along the 
rotation axis,
deviating from the static dipole configuration.
Thus, the collision angles near the light cylinder,
and hence $\sigma_2$ decreases with decreasing $\alpha$.
The explicit value of $\epsilon$ can be computed only numerically,
solving the photon specific intensity from infrared to $\gamma$-ray
energies in the three-dimensional pulsar magnetosphere.

The last factor, $l_1 l_2$, in the left-hand side of
equation~(\ref{eq:S_master}) is given by
\begin{equation} 
  l_1 l_2 = r_0{}^2 \cos^2\theta_0
           (\tan\theta_0-\tan\theta_1)
           (\tan\theta_2-\tan\theta_0)
  \label{eq:l1l2}
\end{equation}
Thus, equation~(\ref{eq:S_master}) gives
\begin{eqnarray}
  &&
  \frac{e E_\parallel}{h\nu_{\rm c}}
  \frac{L_{\rm X}/c}{2.82 kT}
  \epsilon \sqrt{1-\mu_1} \sigma_{\rm T}
  \nonumber\\
  && \times
  \cos^2\theta_0
           (\tan\theta_0-\tan\theta_1)
           (\tan\theta_2-\tan\theta_0)
  = 1,
  \label{eq:S_master2}
\end{eqnarray}
where the $r_0$ dependence vanishes.
Substituting equations~(\ref{eq:app_Ell}),
(\ref{eq:app_Lf}), (\ref{eq:app_Eg}) into (\ref{eq:S_master2}),
we can solve $h_{\rm m}$ as a function of
$L_{\rm X}/kT$, $P$, and $\mu$.

To describe the evolution of $P=P(t)=2\pi/\Omega(t)$,
we adopt in this paper
\begin{equation}
  -I \Omega \dot{\Omega}= C \frac{\mu^2 \Omega^4}{c^3}
\end{equation}
where $C=(2/3)\sin^2\alpha$ for a magnetic dipole braking,
while $C=1+\sin^2\alpha$ for a force-free braking \citep{spit06}.
Assuming a magnetic dipole braking,
we obtain
\begin{equation}
  P= 39.2 \mbox{ms} \mu_{30} I_{45}^{-1/2} (t/10^3 \mbox{years})^{1/2},
\end{equation}
where $\mu_{30} \equiv \mu / (10^{30}\,\mbox{G cm}^3)$
and $I_{45} \equiv I/(10^{45}\,\mbox{g cm}^2)$.
Thus, if we specify a cooling scenario, $T=T(t)$,
equation~(\ref{eq:S_master2}) gives
$h_{\rm m}$ as a function of $t$.
Note that the $\alpha$ dependence of the spin-down law
is not essential for the present purpose;
thus, $C=2/3$ is simply adopted.
Once $h_{\rm m}=h_{\rm m}(t)$ is obtained,
equation~(\ref{eq:Lg_0}) readily gives 
$L_\gamma$ as a function of $t$,
and hence of $L_{\rm spin}$.
It is worth noting that the heated polar-cap emission
is relatively weak compared to the cooling NS emission,
except for millisecond or middle-aged pulsars.

Let us now consider the cooling scenario.
Since the mass of PSR~J1614-2230 is precisely measured to be
$1.97 M_\odot$ (i.e., $1.97$ solar masses),
and since other three NSs 
(4U1700-377, B1957+20, and J1748-2021B) \citep{latti01} 
are supposed to be heavier than $2.0 M_\odot$,
we exclude the equation of state (EOS) that gives smaller
maximum mass than $1.95 M_\odot$.
That is,
we do not consider the fast cooling scenario due to
direct Urca process in a hyperon-mixed, pion-condensed,
kaon-condensed, or quark-deconfined core,
which gives softer EOS and hence a smaller maximum mass.
Even without exotic matters, the direct Urca process may also
become important in the core of a NS
with the mass that is slightly less than the maximum mass.
However, in the present paper,
we exclude such relatively rare cases
and adopt the canonical value, $1.4 M_\odot$, as the NS mass.

On these ground, we adopt the minimal cooling scenario \citep{pag04},
which has no enhanced cooling that could result from
any of the direct Urca processes
and employs the standard EOS, APR EOS \citep{apr98}.
Within the minimal cooling scenario, 
the cooling history of a NS substantially depends
on the composition of the envelope,
which is defined as 
the upper-most layer extending from the photosphere
down to a boundary where the luminosity in the envelope
equals the total surface luminosity of the star.
An envelope contains a strong temperature gradient and
has a thickness around $100$ meters.
We adopt the cooling curves given in \citet{pag04}
and consider the two extreme cases:
light element and heavy element envelopes.
Because of the uncertainty in the modeling of
neutron ${}^1S_0$ (i.e., spin-singlet state) Cooper paring temperature 
(as a function of neutron Fermi momentum)
and proton ${}^1S_0$ paring temperature
(as a function of proton Fermi momentum),
the predicted $L_{\rm X}$ distributes in a \lq band' for each
chemical composition of the envelope.
When a NS is younger than $10^{4.4}\,\mbox{yr}$,
a light element envelope 
(contaminated by e.g., H, He, C, or O with masses exceeding
 $10^{-6} M_\odot$) 
has a higher surface temperature, 
and hence a greater luminosity, $L_{\rm X}$,
compared to a heavy element envelope
(contaminated by light elements with masses less than 
 $10^{-16} M_\odot$),
because the heat transport becomes more efficient in the former.
As the NS ages, a star with a light element envelope
quickly loses its internal energy via photon emission;
as a result, after $10^{4.6}\,\mbox{yr}$,
it become less luminous than that with a heavy element envelope.
A realistic cooling curve will be located 
between these two extreme cases, 
depending on the actual composition of the NS envelope.

\begin{figure}
 \epsscale{1.0}
 \plotone{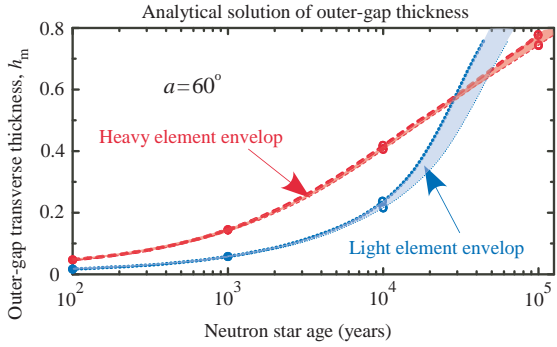} 
\caption{
Analytically solved gap thickness.
The (blue) dotted and (red) dashed curves represent
$h_{\rm m}=h_{\rm m}(t)$ 
for a light and a heavy element envelopes, respectively. 
Uncertainties due to nucleon Cooper paring models,
are represented by the blue and red shaded \lq bands'.
Magnetic inclination angle is assumed to be $\alpha=60^\circ$.
\label{fig:hm_evol}
}
\end{figure}

We present the solved $h_{\rm m}$ for a light and a heavy
element envelope in figure~\ref{fig:hm_evol},
adopting $\mu_{30}=3.2$,
which gives the magnetic field strength of 
$4.1 \times 10^{12}$~G at the magnetic pole,
where $r_\ast=11.6$~km is used.
It follows that the gap becomes thinner for
a light element case 
(between the thin and thick dotted curves, blue shaded region)
than for the heavy element cases
(between the thin and thick dashed curves, red shaded region).
This is because the more luminous photon field of 
a light element envelope leads to a copious pair production,
which prevents the gap to expand in the trans-field direction.
As a result, the predicted $L_\gamma$ 
becomes less luminous for a light element envelope
than a heavy one. 

\begin{figure}
 \epsscale{1.0}
 \plotone{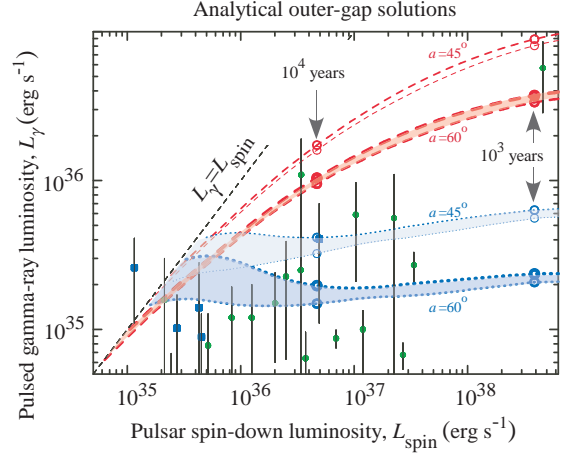} 
\caption{
Analytically solved gap luminosity
with uncertainties due to nucleon Cooper paring models.
The thick and thin dotted (or dashed) curves represent 
the evolution of the 
gap luminosity for a light (or a heavy) element envelope.
Two cases of magnetic inclination angle, 
$\alpha=45^\circ$ and $60^\circ$, are depicted as labeled.
The green filled circles designates the normal gamma-ray pulsars,
while the blue filled squares do those detected 
by the gamma-ray blind search technique.
\label{fig:LgLsp_1}
}
\end{figure}

In figure~\ref{fig:LgLsp_1},
we present the analytical results of $L_\gamma$ versus
$L_{\rm spin}$
as the dotted (or dashed) curve for a light (or a heavy) 
element envelope.
As the pulsar spins down, $L_\gamma$ evolves leftwards.
In this section, for analytical purpose, 
we are not interested in the dependence on the observer's 
viewing angle, $\zeta$. 
Thus, in equation~(\ref{eq:Lg_0}), we simply put
$f_\Omega=1$.
It is interesting to note that $L_\gamma$ little
evolves for a light element envelope,
because of $h_{\rm m} \ll 1$,
as discussed after equation~(\ref{eq:S_th0}).
Precisely speaking, we obtain
$h_{\rm m} \propto t^{0.62}$ and hence 
$L_\gamma \propto L_{\rm spin}{}^{0.07}$
(or $h_{\rm m} \propto t^{0.50}$ and hence 
 $L_\gamma \propto L_{\rm spin}{}^{0.25}$)
for a light (or a heavy) element envelope.
That is, although $h_{\rm m}$ is smaller,
$h_{\rm m}$ increases more rapidly in a light element case
than in a heavy element case,
which enables a constant $L_\gamma$ for a light element envelope.
At later stage, $t>10^4$ years, 
both $h_{\rm m}$ and $L_\gamma$ increase with decreasing $L_{\rm spin}$
for a light element envelope,
because $L_{\rm X}/(kT)$, and hence the pair production rate, 
rapidly decreases with increasing $t$.
On the other hand, for a heavy element envelope,
their greater $h_{\rm m}$ results in a monotonically decreasing
$L_\gamma$ with decreasing $L_{\rm spin}$.

\section{Numerical examination of outer-gap electrodynamics}
\label{sec:numerical}
Let us develop the analytical examination
and look deeper into a self-consistent solution by 
a numerical method.
To this end, we adopt the modern outer-gap model \citep{hiro11a}
and solve the set of Maxwell and Boltzmann equations
self-consistently
and compute $E_\parallel$, distribution functions of $e^\pm$'s,
and the photon specific intensity
at each point in the three-dimensional pulsar magnetosphere.
We consider not only the whole-surface, cooling NS emission
but also the heated polar-cap emission as the photon source of
photon-photon pair production in the numerical analysis.
The former emission component is given as a function of the 
pulsar age from the minimum cooling scenario, 
in the same manner as in the analytical examination,
while the latter emission component is solved 
consistently with the energy flux of the $e^-$'s falling on to
the pulsar polar-cap surface.
The method described below was also used in recent
theoretical computation of the Crab pulsar's $\gamma$-ray
emissions \citep{ale11a,ale11b}.

Let us present the basic equations that describe
a pulsar outer-magnetospheric accelerator, 
extending the method first proposed for black-hole 
magnetospheres \citep{bes92,hir98}.
Around a rotating NS,
the background geometry is 
described by the space-time metric \citep{len18}
\begin{equation}
  ds^2= g_{tt} dt^2 + 2g_{t\varphi}dtd\varphi
       +g_{rr} dr^2 + g_{\theta\theta} d\theta^2
       +g_{\varphi\varphi} d\varphi^2,
  \label{eq:metric_1}
\end{equation}
where
\begin{equation}
  g_{tt}       
    \equiv \left( 1-\frac{r_{\rm g}}{r}\right) c^2, \,
  g_{t\varphi} 
    \equiv ac\frac{r_{\rm g}}{r}\sin^2\theta,
  \label{eq:metric_2}
\end{equation}
\begin{equation}
  g_{rr}   \equiv -\left( 1-\frac{r_{\rm g}}{r}\right)^{-1}, \,
  g_{\theta\theta}   \equiv -r^2, \,
  g_{\varphi\varphi} \equiv -r^2 \sin^2\theta;
  \label{eq:metric_3}
\end{equation}
$M$ denotes the NS mass, 
$r_{\rm g}\equiv 2GM/c^2$ the Schwarzschild radius,
and $a \equiv I\Omega/(Mc)$  the stellar angular momentum.
At radial coordinate $r$, the inertial frame is dragged
at angular frequency 
$ \omega \equiv -g_{t\varphi}/g_{\varphi\varphi}
  = 0.15 \Omega I_{45} r_6{}^{-3} $,
where $I_{45} \equiv I/10^{45} \mbox{ erg cm}^2$, and 
$r_6 \equiv r_\ast/10\,\mbox{km}$.

First, let us derive the Poisson equation for the electrostatic 
potential under the space-time geometry described just above.
Let us consider the Gauss's law,
\begin{equation}
  \nabla_\mu F^{t\mu}
  = \frac{1}{\sqrt{-g}}
    \partial_\mu \left[ \frac{\sqrt{-g}}{\rho_{\rm w}^2}
                       g^{\mu\nu}(-g_{\varphi\varphi}F_{t\nu}
                               +g_{t\varphi}F_{\varphi\nu})
               \right]
  = \frac{4\pi}{c^2} \rho,
  \label{eq:Poisson_1}
\end{equation}
where $\nabla$ denotes the covariant derivative,
the Greek indices run over $t$, $r$, $\theta$, $\varphi$;
$\sqrt{-g}= \sqrt{g_{rr}g_{\theta\theta}\rho_{\rm w}^2}=cr^2\sin\theta$ and
$\rho_{\rm w}^2 \equiv g_{t\varphi}^2-g_{tt}g_{\varphi\varphi}$,
$\rho$ the real charge density.
The electromagnetic fields observed by a distant static
observer are given by~\citep{Came86a,Came86b}
$ E_r=F_{rt}, \, E_\theta=F_{\theta t}, \, E_\varphi=F_{\varphi t}$,
$ B^r= (g_{tt}+g_{t\varphi}\Omega) F_{\theta\varphi}/\sqrt{-g}, \,
  B^\theta= (g_{tt}+g_{t\varphi}\Omega) F_{\varphi r}/\sqrt{-g}, \,
  B_\varphi= -\rho_{\rm w}^2 F_{r \theta}/\sqrt{-g}$,
where $F_{\mu\nu} \equiv \partial_\mu A_\nu-\partial_\nu A_\mu$ 
and $A_\mu$ denotes the vector potential.

We assume that the electromagnetic fields are unchanged
in the co-rotating frame. 
In this case, it is convenient to introduce 
the non-corotational potential $\Psi$ that satisfies
\begin{equation}
  F_{\mu t}+\Omega F_{\mu \varphi}
  = -\partial_\mu \Psi(r,\theta,\varphi-\Omega t),
  \label{eq:def_Psi}
\end{equation}
where $\mu= t,r,\theta,\varphi$.
If $F_{A t}+\Omega F_{A \varphi}=0$ holds for $A=r$ and $\theta$,
the magnetic field rigidly rotates with angular frequency $\Omega$.
Imposing that the NS surface a perfect conductor,
$F_{\theta t}+\Omega F_{\theta\varphi}=0$,  
we find that the NS surface becomes equi-potential,
$\partial_\theta \Psi= \partial_t \Psi +\Omega \partial_\varphi \Psi=0$.
However, in a particle acceleration region,
the magnetic field does not rigidly rotate, because
$F_{A t}+\Omega F_{A \varphi}$ deviates from $0$.
The deviation is expressed in terms of $\Psi$, which gives
the strength of the acceleration electric field, 
\begin{equation}
  E_\parallel \equiv \frac{\mbox{\boldmath$B$}}{B}
              \cdot \mbox{\boldmath$E$}
       = \frac{B^i}{B}(F_{it}+\Omega F_{i\varphi})
       = \frac{\mbox{\boldmath$B$}}{B}
              \cdot (-\nabla\Psi),
  \label{eq:def_Ell}
\end{equation}
which is measured by a distant static observer, 
where the Latin index $i$ runs over spatial coordinates
$r$, $\theta$, $\varphi$.

Substituting equation~(\ref{eq:def_Psi}) into (\ref{eq:Poisson_1}),
we obtain the Poisson equation for $\Psi$,
\begin{equation}
  -\frac{c^2}{\sqrt{-g}}
   \partial_\mu 
      \left( \frac{\sqrt{-g}}{\rho_{\rm w}^2}
             g^{\mu\nu} g_{\varphi\varphi}
             \partial_\nu \Psi
      \right)
  = 4\pi(\rho-\rho_{\rm GJ}),
  \label{eq:Poisson_2}
\end{equation}
where 
\begin{equation}
  \rho_{\rm GJ} \equiv 
      \frac{c^2}{4\pi\sqrt{-g}}
      \partial_\mu \left[ \frac{\sqrt{-g}}{\rho_{\rm w}^2}
                         g^{\mu\nu} g_{\varphi\varphi}
                         (\Omega-\omega) F_{\varphi\nu}
                 \right]
  \label{eq:def_GJ}
\end{equation}
denotes the general relativistic Goldreich-Julian charge density.
If $\rho=\rho_{\rm GJ}$ holds everywhere,
$E_\parallel$ vanishes in the entire region,
provided that the boundaries are equi-potential.
However, if $\rho$ deviates from $\rho_{\rm GJ}$ in any region,
$E_\parallel$ appears around the region with
differentially rotating magnetic field lines.
In the higher altitudes, $r \gg r_{\rm g}$, 
equation~(\ref{eq:def_GJ}) reduces to the
special-relativistic expression~\citep{gol69,mes71},
\begin{equation}
  \rho_{\rm GJ} 
  \equiv -\frac{\mbox{\boldmath$\Omega$}\cdot\mbox{\boldmath$B$}}
               {2\pi c}
         +\frac{(\mbox{\boldmath$\Omega$}\times\mbox{\boldmath$r$})\cdot
                (\nabla\times\mbox{\boldmath$B$})}
               {4\pi c},
  \label{eq:def_rhoGJ_1}
\end{equation}
which is commonly used.

Instead of ($r$,$\theta$,$\varphi$),
we adopt the magnetic coordinates
($s$,$\theta_\ast$,$\varphi_\ast$),
where $s$ denotes the distance along a magnetic field line,
$\theta_\ast$ and $\varphi_\ast$ represents the magnetic 
co-latitude and the magnetic azimuthal angle, respectively,
of the point where the field line intersects the NS surface.
Defining that $\theta_\ast=0$ corresponds to the magnetic axis
and that $\varphi_\ast=0$ to the latitudinal plane on which both 
the rotation and the magnetic axes reside,
we obtain the Poisson equation~\citep{hir06a,hir06b},
\begin{eqnarray}
  &&
  -\frac{c^2 g_{\varphi\varphi}}{\rho_{\rm w}^2}
  \left( g^{ss}\partial_s^2 
        +g^{\theta_\ast \theta_\ast} \partial_{\theta_\ast}^2
        +g^{\varphi_\ast \varphi_\ast}
            \partial_{\varphi_\ast}^2
  \right. \nonumber\\
  && \hspace*{1.2 truecm}
  \left.
        +2g^{s\theta_\ast} \partial_s \partial_{\theta_\ast}
        +2g^{\theta_\ast \varphi_\ast}
            \partial_{\theta_\ast} \partial_{\varphi_\ast}
        +2g^{\varphi_\ast s}
            \partial_{\varphi_\ast} \partial_s
  \right) \Psi
  \nonumber\\
  && -\left( A^s \partial_s
            +A^{\theta_\ast} \partial_{\theta_\ast}
            +A^{\varphi_\ast} \partial_{\varphi_\ast}
      \right) \Psi
  = 4\pi (\rho-\rho_{\rm GJ}),
  \label{eq:BASIC_1}
\end{eqnarray}
where
\begin{eqnarray}
  g^{i'j'} 
  && = g^{\mu\nu}\frac{\partial x^{i'}}{\partial x^\mu}
                \frac{\partial x^{j'}}{\partial x^\nu}
     = g^{rr}\frac{\partial x^{i'}}{\partial r}
            \frac{\partial x^{j'}}{\partial r}
      +g^{\theta\theta}\frac{\partial x^{i'}}{\partial \theta}
                     \frac{\partial x^{j'}}{\partial \theta}
  \nonumber\\
  && \hspace{0.5 truecm}
      -\frac{k_0}{\rho_{\rm w}^2}
       \frac{\partial x^{i'}}{\partial\varphi}
       \frac{\partial x^{j'}}{\partial\varphi},
  \label{eq:def_mag1}
\end{eqnarray}
\begin{eqnarray}
  A^{i'} 
  &\equiv&
  \frac{c^2}{\sqrt{-g}}
   \left\{ \partial_r \left[ \frac{g_{\varphi\varphi}}{\rho_{\rm w}^2}
                             \sqrt{-g}g^{rr}
                             \frac{\partial x^{i'}}{\partial r}
                      \right]
   \right. \nonumber\\
  && \hspace*{-0.4 truecm}
   \left. +\partial_\theta
                      \left[ \frac{g_{\varphi\varphi}}{\rho_{\rm w}^2}
                             \sqrt{-g}g^{\theta\theta}
                             \frac{\partial x^{i'}}
                                  {\partial \theta}
                      \right]
    \right\}
   -\frac{c^2 g_{\varphi\varphi}}{\rho_{\rm w}^2}
    \frac{k_0}{\rho_{\rm w}^2}
    \frac{\partial^2 x^{i'}}{\partial\varphi^2};
  \label{eq:def_mag2}
\end{eqnarray}
the coordinate variables are 
$x^1=r$, $x^2=\theta$, $x^3=\varphi$,
$x^{1'}=s$, $x^{2'}=\theta_\ast$, and $x^{3'}=\varphi_\ast$.
Note that this formalism is applicable to arbitrary magnetic field 
configurations,
and also that equation~(\ref{eq:def_Ell}) gives 
$E_\parallel = -(\partial\Psi/\partial s)_{\theta_\ast,\varphi_\ast}$.
Causality requires that a plasmas can co-rotate 
with the magnetic field only in the region that satisfies 
$k_0 \equiv g_{tt}+2g_{t\varphi}\Omega+g_{\varphi\varphi}\Omega^2 > 0$
\citep{zna77,tak90}.
That is, the concept of the light cylinder is generalized 
into the light surface on which $k_0$ vanishes.
The effects of magnetic field expansion ~\citep{sch78,MT92} 
are contained in the coefficients of the trans-field derivatives,
$g^{\theta_\ast\theta_\ast}$,
$g^{\theta_\ast\varphi_\ast}$,
$g^{\varphi_\ast\varphi_\ast}$.
In what follows, we adopt the vacuum, rotating dipole 
solution~\citep{cheng00} to describe the magnetic field configuration.

Second, let us consider the particle Boltzmann equations.
At time $t$, position $\mbox{\boldmath$r$}$, 
and momentum $\mbox{\boldmath$p$}$, they become,
\begin{equation}
  \frac{\partial{N_\pm}}{\partial t}
    + \mbox{\boldmath$v$}
      \cdot \mbox{\boldmath$\nabla$} N_\pm
    + \left( q\mbox{\boldmath$E$}
             +\frac{\mbox{\boldmath$v$}}{c}
              \times\mbox{\boldmath$B$}
      \right) \cdot 
      \frac{\partial N_\pm}{\partial \mbox{\boldmath$p$}}
  = S_\pm (t,\mbox{\boldmath$r$},\mbox{\boldmath$p$}),
  \label{eq:boltz_1}
\end{equation}
where 
$N_+$ (or $N_-$) denotes the positronic (or electronic)
distribution function; 
$\mbox{\boldmath$v$} \equiv $\mbox{\boldmath$p$}$/(m_{\rm e}\gamma)$,
$m_{\rm e}$ refers to the rest mass of the electron,
and $q$ the charge on the particle.
The Lorentz factor is given by 
$\gamma \equiv 1/\sqrt{1-(\vert \mbox{\boldmath$v$} \vert /c)^2}$.
The collision term $S_+$ (or $S_-$) consists of the terms 
that represent the appearing and disappearing 
rates of positrons (or electrons) at $\mbox{\boldmath$r$}$
and $\mbox{\boldmath$p$}$ per unit time per unit phase-space volume.
Since we are dealing with high-energy phenomena,
we consider wave frequencies that are much greater than the
plasma frequency 
and neglect the collective effects.

It is noteworthy that the particle flux per magnetic flux tube
is conserved along the flow line
if there is no particle creation or annihilation.
Thus, it is convenient to normalize the particle distribution functions
by the Goldreich-Julian number density such that
$n_\pm = \langle N_\pm \rangle / (\Omega B/2\pi ce)$,
where $\langle \rangle$ denotes 
that the quantity is averaged in a gyration.
Imposing a stationary condition 
$\partial/\partial t
  + \Omega \partial/\partial \phi = 0
  \label{eq:stationary}
$
in the co-rotating frame,
we can reduce the particle Boltzmann equations into
\citep{hir03}
\begin{equation}
  c\cos\chi  \frac{\partial n_\pm}{\partial s}
  +\dot{p}   \frac{\partial n_\pm}{\partial p}
  +\dot{\chi}\frac{\partial n_\pm}{\partial \chi}
  = S_{\rm IC} +S_{\rm p},
 \label{eq:BASIC_2}
\end{equation}
where the upper and lower signs correspond to the
positrons (with charge $q=+e$) and electrons ($q=-e$), respectively, 
$\chi$ denotes the pitch angle of gyrating particles, and
$p \equiv \vert\mbox{\boldmath$p$}\vert=m_{\rm e}c\sqrt{\gamma^2-1}$.
Since pair annihilation is negligible in a pulsar magnetosphere,
$S_\pm$ consists of the IC scattering term, $S_{\rm IC}$, and 
the pair creation term, $S_{\rm p}$, in the right-hand side.
When a particle emits a photon via synchro-curvature process,
the energy loss ($\sim$~GeV) is small compared to the
particle energy ($\sim 10$~TeV);
thus, it is convenient to include the back reaction of 
the synchro-curvature emission
as a friction term in the left-hand side~\citep{hir06a}.
In this case, 
the characteristics of equation~(\ref{eq:BASIC_2}) 
in the phase space are given by 
\begin{equation}
  \dot{p} \equiv q E_\parallel \cos\chi -\frac{P_{\rm SC}}{c},
  \label{eq:char1}
\end{equation}
\begin{equation}
  \dot{\chi} \equiv -\frac{q E_\parallel \sin\chi}{p}
                    +c\frac{\partial(\ln B^{1/2})}{\partial s}
                     \sin\chi,
  \label{eq:char2}  
\end{equation}
where the synchro-curvature radiation force, $P_{\rm SC}/c$ 
\citep{cheng96,zhang97}, is included as the friction;
the particle position $s$ is related with time $t$ by 
$\dot{s}= ds/dt= c \cos\chi$.
For outward- (or inward-) migrating particles,
$\cos\chi>0$ (or $\cos\chi<0$).
If $E_\parallel=0$, 
particles will be reflected by magnetic mirrors,
which are expressed by the second term in the
right-hand side of equation~(\ref{eq:char2}.
If we integrate $n_\pm$ over $p$ and $\chi$, 
and multiply the local GJ number density, $\Omega B/2\pi ce$,
we obtain the spatial number density of particles.
Therefore, we can express the real charge density $\rho$ as
\begin{equation}
  \rho
  = \frac{\Omega B}{2\pi c}
    \int\!\!\!\!\int ( n_+ - n_- ) d\gamma d\chi
   +\rho_{\rm ion},
  \label{eq:def_rhoe}
\end{equation}
where $n_\pm$ are a function of 
$s$, $\theta_\ast$, $\varphi_\ast$, $\gamma$, $\chi$;
$\rho_{\rm ion}$ refers to the charge density of ions,
which can be drawn from the NS surface
as a space-charge-limited flow (SCLF) 
by a positive $E_\parallel$ \citep{hir06a}.


In equation~(\ref{eq:BASIC_2}), 
the collision terms are expressed as
\begin{eqnarray}
  S_{\rm IC} 
  &\equiv&
  -\sum_n^{\epsilon_{n-1}<\gamma}
        \eta_{\rm IC}^\gamma (\langle\epsilon_n\rangle,\gamma,\mu_{\rm c}) 
        \int\!\!\!\!\int n_\pm d\gamma d\chi,
  \nonumber\\
  && 
  +\sum_{i} 
        \eta_{\rm IC}^{\rm e} (\gamma_i, \gamma, \mu_{\rm c})
        \int\!\!\!\!\int n_\pm d\gamma d\chi,
   \label{eq:src_1a}
\end{eqnarray}
and
\begin{equation}
  S_{\rm p} \equiv
  \frac{2\pi ce}{\Omega B}
  \int (\alpha_{\gamma B}+\alpha_{\gamma\gamma}) 
  \int \frac{I_\nu}{h\nu} d\tilde\omega d\nu,
  \label{eq:src_1b}
\end{equation}
where $\langle\epsilon_n\rangle=(\epsilon_{n-1}+\epsilon_n)/2$
represents the typical photon energy within the interval
[$m_{\rm e}c^2\epsilon_{n-1}$,$m_{\rm e}c^2\epsilon_n$],
$\mu_{\rm c}$ the cosine of the collision angle 
between the particles and the soft photons,
$I_\nu$ the specific intensity of the radiation field,
$\tilde\omega$ the solid angle into which the photons are propagating.
To compute $S_{\rm p}$ at each point,
we have to integrate $I_\nu/(h\nu)$ in all directions
to calculate the differential photon number flux, 
$\int I_\nu/(h\nu) d\tilde\omega$.
The absorption coefficients for magnetic and photon-photon 
pair production processes, are given by
\begin{eqnarray}
  \alpha_{\gamma B}
  &=& \frac{e^2 / \hbar c}{4.4}
      \frac{m_{\rm e}c}{\hbar} {B'}_\perp
      \exp\left(-\frac{8}{3} \frac{1}{\epsilon_\gamma {B'}_\perp}
          \right)
  \nonumber\\
  && \quad
      \times
      \delta(\gamma-\gamma_0) \delta(\chi-\chi_0),
  \label{eq:def_alf1}\\
  \alpha_{\gamma\gamma}
  &=& \frac{1-\cos\theta_{\rm c}}{2}
      \int_{\epsilon_{\rm th}}^\infty
        d\epsilon_{\rm s} \frac{dF_{\rm s}}{d\epsilon_{\rm s}}
        \frac{\partial^2 \sigma_{\gamma\gamma}}{d\gamma d\chi},
  \label{eq:def_alf2}
\end{eqnarray}
where ${B'}_\perp \equiv B\sin\theta_{\rm c}/B_{\rm cr}$;
explicit expression of 
$\eta_{\rm IC}^\gamma$ and $\partial^2 \sigma_{\gamma\gamma}/d\gamma d\chi$
are given in the literature (e.g., \citet{hir03}). 
In equation~(\ref{eq:def_alf1}),
${B'}_\perp$ contains the collision angle, $\theta_{\rm c}$, 
between the photon and the magnetic field.
In equation~(\ref{eq:def_alf2}),
$\theta_{\rm c}$ does that between the two photons.
%
The IC redistribution function 
$\eta_{\rm IC}^\gamma (\epsilon_\gamma,\gamma,\mu_{\rm c})$ 
represents the probability
that a particle with Lorentz factor $\gamma$ up-scatters photons 
into energies between 
$m_{\rm e}c^2 \epsilon_\gamma$ and 
$m_{\rm e}c^2(\epsilon_\gamma+d\epsilon_\gamma)$
per unit time when the 
collision angle is $\cos^{-1}\mu_{\rm c}$,
where $\epsilon_\gamma=h\nu/(m_{\rm e}c^2)$
refers to the dimensionless photon energy.
On the other hand, 
$\eta_{\rm IC}^{\rm e} (\gamma_i,\gamma_f,\mu_{\rm c})$ represents
the probability for a particle to change its Lorentz factor from
$\gamma_i$ to $\gamma_f$ in a single scattering.
We thus obtain
$ \eta_{\rm IC}^{\rm e}(\gamma_i,\gamma_f,\mu_{\rm c}) 
  = \eta_{\rm IC}^\gamma (\gamma_i-\gamma_f,\gamma_i,\mu_{\rm c})$
by energy conservation.

Third, let us consider 
the radiative transfer equation.
The variation of specific intensity, $I_\nu$, 
along a ray is described by the radiative transfer equation,
\begin{equation}
  \frac{d I_\nu}{dl}= -\alpha_\nu I_\nu + j_\nu,
  \label{eq:RTE}
\end{equation}
where $l$ refers to the distance along the ray,
$\alpha_\nu$ and $j_\nu$ 
the absorption and emission coefficients, respectively.
Both $\alpha_\nu$ and $j_\nu$ are 
a function of $l$, photon energy $E_\gamma$, and 
propagation direction ($k^\theta$,$k^\varphi$),
where $k^\mu$ denotes the photon momentum four vector 
($\mu=t,r,\theta,\varphi$).

Equation~(\ref{eq:RTE}) can be solved 
if we specify the photon propagation in the curved space time.
The evolution of momentum and position of a photon 
is described by the Hamilton-Jacobi equations,
\begin{equation}
 c\frac{dk_r     }{dl}=-\frac{\partial k_t}{\partial r},
  \quad
 c\frac{dk_\theta}{dl}=-\frac{\partial k_t}{\partial \theta},
  \label{eq:HJ_2a}
\end{equation}
\begin{equation}
 c\frac{dr      }{dl}=\frac{\partial k_t}{\partial k_r      },
  \quad
 c\frac{d\theta }{dl}=\frac{\partial k_t}{\partial k_\theta }.
  \label{eq:HJ_2b}
\end{equation}
Since the metric (eqs.~[\ref{eq:metric_1}]--[\ref{eq:metric_3}])
is stationary and axisymmetric, the photon energy at infinity $k_t$ 
and the azimuthal wave number $-k_\varphi$ are conserved along the ray.
When a particle is rotating with angular velocity $\dot\varphi$
and emit a photon with energy $E_{\rm local}$, 
$k_t$ and $-k_\varphi$ are related to these quantities by 
the redshift relation,
$E_{\rm local}= (dt/d\tau)(k_t+k_\varphi \dot\varphi)$,
where $dt/d\tau$ is solved from
the definition of the proper time,
$(dt/d\tau)^2(g_{tt}+2g_{t\varphi}\dot{\varphi}
 +g_{\varphi\varphi}\dot{\varphi}^2)=1$.
The dispersion relation $k^\mu k_\mu=0$,
which is quadratic to $k_\mu$'s ($\mu=t,r,\theta,\varphi$),
gives Hamiltonian $k_t$ 
in terms of $r$, $\theta$, $k_r$, $k_\theta$, and $k_\varphi$.
Thus, we have to solve the set of four ordinary differential 
equations~(\ref{eq:HJ_2a}) and (\ref{eq:HJ_2b})
for $k_r$, $k_\theta$, $r$, and $\theta$.
When photons are emitted, 
they are highly beamed along the particle's motion;
thus, the initial conditions of ($k^r$,$k^\theta$,$k^\varphi$)
are given by the instantaneous particle's velocity 
measured by a distant static observer.
It may be helpful to give the initial conditions of ray tracing
in the limit $r \gg r_{\rm g}$ and $\gamma \gg 1$,
because photons are emitted mostly in the outer magnetosphere.
In this limit,
the orthonormal components of the $e^+$'s and $e^-$'s
instantaneous velocity 
are given by~\citep{mes85,Came86a,Came86b} 
\begin{equation}
  \frac{v^r}{c}
  = f_{\rm v} \frac{B^r}{c}, \quad
  \frac{v^{\hat\theta}}{c}
  = f_{\rm v} \frac{B^{\hat\theta}}{c}, \quad
  \frac{v^{\hat\phi}}{c}
  = f_{\rm v} \frac{B^{\hat\varphi}}{c} + \frac{\varpi}{\varpi_{\rm LC}}
\end{equation}
in the polar coordinates, where
\begin{equation}
  f_{\rm v} 
  \equiv -\frac{\varpi}{\varpi_{\rm LC}}\frac{B^{\hat\varphi}}{B}
         \pm \sqrt{1-\left(\frac{\varpi}{\varpi_{\rm LC}}\right)^2
                     \left(\frac{B_{\rm p}}{B}\right)^2};
  \label{eq:v3Df}
\end{equation}
$B_{\rm p}\equiv \sqrt{(B^r)^2+(B^{\hat\theta})^2}$.
The upper (or the lower) sign of $f_{\rm v}$ (eq.~[\ref{eq:v3Df}]) 
corresponds to the outward (or the inward) particle velocity.
Note that $B^{\hat\varphi}<0$ holds in ordinary situation.

Fourth and finally, let us impose appropriate boundary
conditions to solve the set of 
Maxwell (i.e., Poisson) and Boltzmann equations.
We start with considering the boundary conditions for 
the elliptic type equation~(\ref{eq:Poisson_2}).
We define that the {\it inner} boundary, $s=0$,
coincides with the NS surface, on which we put $\Psi=0$ for convenience.
The {\it outer} boundary, $s=s_{\rm out}(\theta_\ast,\varphi_\ast)$,
is defined as the place where
$E_\parallel$ changes sign near the light cylinder.
Its location is solved self-consistently as a free-boundary problem
and appears near the place where 
$\partial(\rho_{\rm GJ}/B)/\partial s$ vanishes due to the flaring up of the
field lines towards the rotation axis 
(eq.~[68] of~\citet{hir06a}).
At each $\varphi_\ast$, the {\it lower} boundary 
$\theta_\ast=\theta_\ast^{\rm max}(\varphi_\ast)$ 
is assumed to coincide with the last open field line, 
which is defined by the condition that 
$ \sqrt{-g_{rr}}B^r \sin\theta 
 +\sqrt{-g_{\theta\theta}}B^\theta \cos\theta=0$ is satisfied at the 
light cylinder.
We can compute the potential drop along each field line,
$\Delta\Psi(\theta_\ast,\varphi_\ast) 
 \equiv \Psi(s=0)-\Psi(s=s_{\rm out})
 = -\Psi(s=s_{\rm out})$,
by integrating $E_\parallel$ along the field line.
The maximum value of $\Delta\Psi(\theta_\ast,\varphi_\ast)$ 
is referred to as $\Delta\Psi_{\rm max}$.

Let us consider the {\it upper} boundary, 
$\theta_\ast^{\rm min}= \theta_\ast^{\rm min}(s,\varphi_\ast)$.
At each ($s$,$\varphi_\ast$),
$E_\parallel$ vanishes on the last-open field line,
$\theta_\ast=\theta_\ast^{\rm max}$
(i. e., at the lower boundary),
and increases with decreasing $\theta_\ast$
in the latitudinal direction
(i. e., towards the magnetic axis).
The acceleration field $E_\parallel$ 
peaks around the middle-latitudes,
$\theta_\ast \approx 
 [\theta_\ast^{\rm max}+\theta_\ast^{\rm min}]/2
 = (1-h_{\rm m}/2) \theta_\ast^{\rm max}$,
and turns to decrease towards the upper boundary.
At some co-latitude,
$E_\parallel$ eventually decreases below 
$0.01 \Delta\Psi_{\rm max}/\varpi_{\rm LC}$;
we define this co-latitude as the gap upper boundary,
$\theta_\ast^{\rm min}= \theta_\ast^{\rm min}(s,\varphi_\ast)$.
To specify the magnetic field line at each $\varphi_\ast$,
it is convenient to introduce the dimensionless co-latitude,
\begin{equation}
  h \equiv 1-\theta_\ast/\theta_\ast^{\rm max}(\varphi_\ast).
  \label{eq:def_h}
\end{equation}
For example, $h=0$ specifies the last-open field line,
and $h=0.1$ does the field line having the foot point on the
polar-cap (PC) surface at $\theta_\ast=0.9\theta_\ast^{\rm max}$ 
at each magnetic azimuthal angle $\varphi_\ast$.
At the upper boundary, we obtain
\begin{equation}
  h= h_{\rm m}(s,\varphi_\ast)
   \equiv 1-\theta_\ast^{\rm min}(s,\varphi_\ast)
           /\theta_\ast^{\rm max}(\varphi_\ast).
  \label{eq:def_hm}
\end{equation}
To solve the Poisson equation, we put $\Psi=0$ 
at $h=h_{\rm m}$, or equivalently, 
at $\theta_\ast=\theta_\ast^{\rm min}(s,\varphi_\ast)$.
If $h_{\rm m} \ll 1$, the gap becomes  thin, whereas
$h_{\rm m}\sim 1$ indicates that the gap is threaded by 
most of the open-field lines.  
Note that the open field lines cross the PC surface
at magnetic co-latitudes
$\theta_\ast^{\rm min} =(1-h_{\rm m})\theta_\ast^{\rm max}
 < \theta_\ast =(1-h)\theta_\ast{}^{\rm max} < \theta_\ast^{\rm max}$
(i.e., $h_{\rm m}>h>0$),
and that $\theta_\ast=0$ (i.e., $h=1$) corresponds to the magnetic axis.

We also have to consider the boundary conditions for 
the hyperbolic type equations~(\ref{eq:BASIC_2}) and (\ref{eq:RTE}).
(Eq.~[\ref{eq:RTE}] itself is an ordinary differential equation;
 however, it is equivalent to solving the Boltzmann equation of photon
 distribution function.)
We assume that $e^\pm$'s and photons are not injected
into the gap across either the inner or the outer boundaries.
However, if the created current becomes greater than
the GJ value, 
a positive $E_\parallel$ arises at the NS surface to draw
ions from the surface as a SCLF
until $E_\parallel$ almost vanishes at the surface.

To sum up, we solve the set of partial and ordinary 
differential equations
(\ref{eq:Poisson_2}), (\ref{eq:BASIC_2}), and (\ref{eq:RTE})
under the boundary conditions described in the three
foregoing paragraphs.
By this method, we can solve 
the acceleration electric field $E_\parallel$,
particle distribution functions $n_\pm$, and
the photon specific intensity $I_\nu$ (from $h\nu=0.005$~eV to $50$~TeV),
at each position in the three-dimensional magnetosphere of
arbitrary rotation-powered pulsars,
if we specify $P$, $\mu$, $\alpha$, and $kT$,
where the surface temperature $kT$ is necessary to
compute the photon-photon pair production
through the differential flux $dF_{\rm s}/d\epsilon_{\rm s}$
in equation~(\ref{eq:def_alf2}).
We adopt the minimum cooling scenario in the same manner as in 
\S~\ref{sec:analytical}.

\begin{figure}
 \epsscale{1.0}
 \plotone{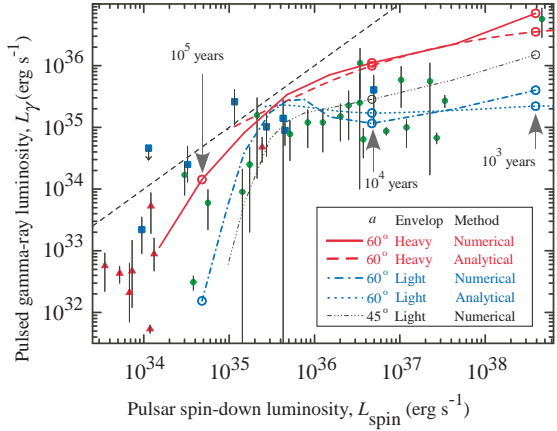} 
\caption{
Same figure as figure~\ref{fig:LgLsp_1}, but 
numerical results are plotted as 
dash-dotted and solid curves for a light and a heavy element envelope,
respectively.
For comparison, a numerical solution for $\alpha=45^\circ$
and a light element envelope,
is also plotted as the dash-dot-dot-dot curve.
To avoid complications,
we adopt the geometrical mean of the 
upper-bound $L_{\rm X}$ and 
lower-bound $L_{\rm X}$ of cooling curves
for each chemical composition,
instead of depicting a \lq band',
which reflects the uncertainties in the nucleon Cooper pairing models.
\label{fig:LgLsp_2}
}
\end{figure}

In figure~\ref{fig:LgLsp_2},
we plot the result of $L_\gamma$ as a function of 
$L_{\rm spin}$
as the dash-dotted (or solid) curve
for a light (or a heavy) element envelope,
where $\mu_{30}=3.2$ is adopted in the same manner as in the
analytical examination.
It follows that these numerical solutions 
are consistent with the analytical ones,
and that $L_\gamma$ decreases slowly until 
$10^{4.5}$ years.
The physical reason why $L_\gamma$ increases 
with decreasing $L_{\rm spin}$ at $t>10^4$~years for a light element
envelope,
is the same as described at the end of \S~\ref{sec:analytical}.
A realistic NS will have an envelope composition
between the two extreme cases, light and heavy elements.
Thus, the actual $L_\gamma$'s will distribute between the 
red solid (or dashed) and the blue dash-dotted (or dotted) curves.
However, after $L_\gamma$ approaches $L_{\rm spin}$
(thin dashed straight line; 
 see \citet{wangR11} for the death line argument),
the outer gap survives only along the limited magnetic field lines
in the trailing side of the rotating magnetosphere
because of a less efficient pair production;
as a result, $L_\gamma$ rapidly decreases with decreasing $L_{\rm spin}$.
For a smaller $\alpha$, even for a light element envelope,
$L_\gamma$ monotonically decreases 
as the dash-dot-dot-dot curve shows,
because the gap is located in the higher altitudes,
and because the less efficient pair production there prevents
the produced electric current to increase with decreasing age around
$t \sim 10^{4.5}$~years.

\section{Exponential cutoff energy}
\label{sec:cutoff}
\begin{figure}
 \epsscale{1.0}
 \plotone{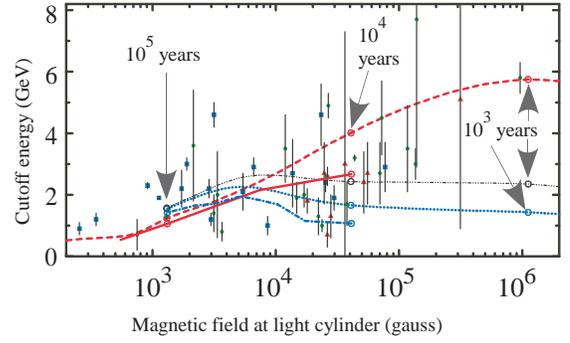} 
\caption{
Cutoff energy of $\gamma$-ray spectrum.
The dashed and dotted curves denote the cutoff energies
obtained analytically for heavy and light element envelopes,
respectively, whereas
the solid and dash-dotted curves do those obtained numerically
for heavy and light element envelopes, respectively.
For these four cases, $\alpha=60^\circ$ is assumed.
For comparison, an analytical solution for $\alpha=45^\circ$
and a light element envelope,
is also plotted as the dash-dot-dot-dot curve.
\label{fig:cutoff}
}
\end{figure}

It is also worth examining the cutoff energy, $E_{\rm cutoff}$, 
of the gamma-ray spectrum.
We plot $E_{\rm cutoff}$ as a function of 
the magnetic-field strength at the light cylinder, $B_{\rm LC}$,
in figure~\ref{fig:cutoff}. 
In the analytical computation 
(dash, dotted, and dash-dot-dot-dot curves),
only the curvature process~\citep{hiro11b} 
from the particles produced and accelerated
in the outer gap, is considered as the photon emission process,
and no photon absorption is considered.
In the numerical computation (solid and dash-dotted curves), 
on the other hand,
synchro-curvature and inverse-Compton processes 
from the particles not only produced and accelerated in the gap
but also cascaded outside the gap,
are considered as the emission processes,
and both the photon-photon and magnetic pair production processes
are taken into account when computing the photon propagation.
Therefore, for very young pulsars,
a strong photon-photon absorption 
(and the subsequent production of lower-energy photons via
 synchrotron and synchrotron-self-Compton processes)
takes place even in the higher altitudes.
As a result, spectrum can no longer be fitted by
power-law with exponential-cutoff functional form.
Thus, for $t<10^4$~years,
the fitted cutoff energies are not plotted for numerical solutions
(i.e., solid and dash-dotted curves).

It follows that the present outer gap model explains
the observed tendency 
that $E_{\rm cutoff}$ increases with increasing $B_{\rm LC}$,
because the Goldreich-Julian charge density (in the gap) increases
with increasing $B_{\rm LC}$.
It also follows that $E_{\rm cutoff}$ is regulated below several GeV,
because copious pair production leads to $h_{\rm m} \ll 1$
for young pulsars (i.e., for strong $B_{\rm LC}$).

\section{Flux correction factor}
\label{sec:flux}
\begin{figure}
 \epsscale{1.0}
 \plotone{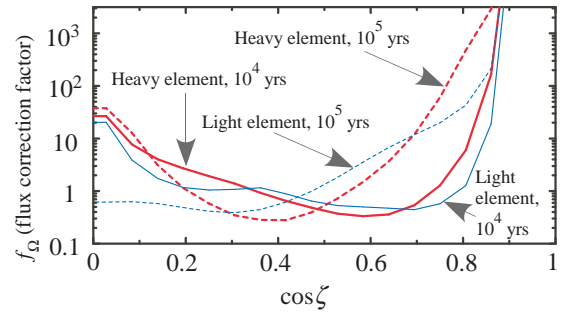} 
\caption{
Flux correction factor, $f_\Omega$, 
as a function of the observer's viewing angle, 
$\zeta$, with respect to the rotation axis.
The (red) thick solid and dashed curves denote 
$f_\Omega$ at pulsar age $10^4$ and $10^5$ years, respectively,
for a heavy element envelope, whereas 
the (blue) thin solid and dashed curves do those
at $10^4$ and $10^5$ years, respectively,
for a light element envelope.
The abscissa denotes the cosine of $\zeta$; 
thus, $\zeta$ distributes randomly along the abscissa
with uniform probability.
For all the cases, $\alpha=60^\circ$ is assumed.
\label{fig:flux}
}
\end{figure}

Finally, let us investigate the flux correction factor,
$f_\Omega$.
To infer the $\gamma$-ray luminosity, 
$L_\gamma=4\pi f_\Omega F_\gamma d^2$,
from the observed flux, $F_\gamma$, 
one has conventionally assumed that the
flux conversion factor is approximately unity, $f_\Omega \approx 1$,
where $d$ denotes the distance to the pulsar.
However, now we can compute $f_\Omega$ explicitly as a function of 
the observer's viewing angle, $\zeta$, 
using the three-dimensional numerical solution.
In figure~\ref{fig:flux},
$f_\Omega(\zeta)$ is depicted for heavy and light element
envelopes at $10^4$ and $10^5$ years.
It follows that the error of $f_\Omega = 1$ is kept 
within a factor of $3$ with a probability greater than $50\%$.
When the pulsar is young, 
$E_\parallel$ is highly screened in the middle and lower altitudes;
as a result, most $\gamma$-rays are emitted from 
the higher altitudes to appear within the 
observer's viewing angle $46^\circ < \zeta < 53^\circ$ 
(i.e., $0.6<\cos\zeta<0.7$).
However, as the pulsar ages,
more $\gamma$-rays are emitted from the middle and lower altitudes, 
resulting in a stronger flux near the rotational equator,
$66^\circ < \zeta < 73^\circ$.

\section{Discussion}
\label{sec:discussion}
To sum up, 
a light element envelope approximately corresponds to the
lower bound of the (observationally inferred) gamma-ray luminosity 
of rotation-powered pulsars,
whereas a heavy element one to the upper bound.
The scatter of the intrinsic gamma-ray luminosity 
is physically determined by
the magnetic inclination angle, $\alpha$, and the envelope composition.
The cutoff energy of the primary curvature emission is
kept below several GeV even for young pulsars,
because the gap trans-field thickness, and hence the acceleration 
electric field, is suppressed by the polarization of the
produced pairs in the lower altitudes.

To convert the observed $\gamma$-ray flux into luminosity, $L_\gamma$,
one has conventionally assumed $f_\Omega=1$.
For example, the error bars of the observational data points 
in figure~\ref{fig:LgLsp_2},
do not contain any uncertainties incurred by $f_\Omega$.
Nevertheless, if $\alpha$ and $\zeta$ can be constrained,
we can estimate $L_\gamma$ more accurately,
by applying the present quantitative outer-gap calculations.
It is noteworthy that $L_\gamma$'s given in 
figures~\ref{fig:LgLsp_1} and \ref{fig:LgLsp_2} 
little depend on the NS magnetic moment, $\mu$.
This is particularly true for a light element case,
which has $h_{\rm m} \ll 1$,
by the reason described after equation~(\ref{eq:S_th0}).
What is more, 
with an additional determination of $d$ 
(e.g., by parallax observations),
we can infer the composition of individual NS envelopes,
by using the constrained flux correction factor,
$f_\Omega$ (fig.~\ref{fig:flux}).
We hope to address such a question as the determination of
$\alpha$ and $\zeta$, and hence $f_\Omega$, for individual pulsars,
by making an \lq atlas' of the pulse profiles and phase-resolved spectra
that are solved from the basic equations
in a wide parameter space of $P$, $\mu$, $T$, $\alpha$, and $\zeta$,
and by comparing the atlas with the observations.



\acknowledgments
The author is indebted to Dr. A.~K. Harding for valuable discussion
on the results.
He also thanks ASPEN Center for Physics 
for providing precious opportunity to debate the main topic
of this letter.
This work is partly supported by 
 the Formosa Program between National Science Council  
 in Taiwan and Consejo Superior de Investigaciones Cientificas
 in Spain administered through grant number 
 NSC100-2923-M-007-001-MY3.

\end{document}